\documentclass[conference]{IEEEtran}
\IEEEoverridecommandlockouts
\usepackage{cite}
\usepackage{amsmath,amssymb,amsfonts}
\usepackage{algorithmic}
\usepackage{graphicx}
\usepackage{textcomp}
\usepackage{xcolor}
\def\BibTeX{{\rm B\kern-.05em{\sc i\kern-.025em b}\kern-.08em
    T\kern-.1667em\lower.7ex\hbox{E}\kern-.125emX}}

\usepackage[labelformat=simple]{subcaption}

\begin{document}

\title{Integrated Sensing and Communication enabled Doppler Frequency Shift Estimation and Compensation

}

\author{\IEEEauthorblockN{Jinzhu Jia\textsuperscript{}, Zhiqing Wei\textsuperscript{}, Ruiyun Zhang\textsuperscript {} and Lin Wang\textsuperscript}\\
	\IEEEauthorblockA{\textsuperscript{} Key
		Laboratory of Universal Wireless Communications, Ministry of Education\\
		School of Information and Communication Engineering\\ Beijing University
		of Posts and Telecommunications, Beijing 100876, China\\
		Email:\{jiajinzhu, weizhiqing, zhangruiyun, wlwl\}@bupt.edu.cn}}

\maketitle

\begin{abstract}
Despite the millimeter wave technology fulfills the low-latency and high data transmission, it will cause severe Doppler Frequency Shift (DFS) for high-speed vehicular network, which tremendously damages the communication performance. In this paper, we propose an Integrated Sensing and Communication (ISAC) enabled DFS estimation and compensation algorithm. Firstly, the DFS is coarsely estimated and compensated using radar detection. Then, the designed preamble sequence is used to accurately estimate and compensate DFS. In addition, an adaptive DFS estimator is designed to reduce the computational complexity. Compared with the traditional DFS estimation algorithm, the improvement of the proposed algorithm is verified in bit error rate and mean square error performance by simulation results.
\end{abstract}

\begin{IEEEkeywords}
Integrated sensing and communication, OFDM, Doppler frequency shift estimation and compensation 
\end{IEEEkeywords}

\section{Introduction}
The autonomous driving system requires not only low latency and highly reliable communication, but also the ability to sensing the surrounding environment. Thus, the Integrated Sensing and Communication (ISAC) is expected to be widely applied in smart transportation, intelligent machine networks, etc \cite{liu fan}. ISAC integrates communication and sensing in a system, which will greatly improve spectrum resource utilization and reduce physical resource complexity\cite{2}.

The real-time broadband transmission and accurate sensing are critical to ensure the safety of autonomous driving. With the advantage of large bandwidth, Millimeter wave is expected to be widely deployed in autonomous driving systems, but it is susceptive to frequency selective fading\cite{3}. Due to the Orthogonal Frequency Division Multiplexing (OFDM) technique divides the frequency band into Multiple orthogonal subchannels, each of which experiences flat fading, it has beneficial anti-fading performance\cite{4}. Nevertheless, the high dynamic DFS in high-speed vehicular networking scenarios will destroy the orthogonality among OFDM subcarriers and cause inter-subcarrier interference, which seriously affects the communication link performance\cite{5}. Lower DFS estimation accuracy may cause larger SNR loss. Therefore, how to achieve real-time accurate estimation and compensation of DFS for high-speed vehicular networking scenarios is an urgent issue.

Compared with cellular communication, the channel of Millimeter wave V2V Communication changes more rapidly, which leads to DFS varying rapidly. For high-speed mobile vehicle networking scenarios, there has been several related works in recent years. Qingpeng Ma et al.\cite{ma} designed an estimator based on hybrid structure training sequences to achieve efficient DFS estimation in high-speed moving vehicle communication. For high speed VANETs (Vehicular Ad Hoc Networks) scenarios, Nyongesa F C et al.\cite{5} proposed a data-assisted Doppler shift estimation method, which utilized a complex exponential expansion matrix to reduce the complexity of the algorithm. The paper \cite{6} proposes a maximum-likelihood Doppler shift estimation method with an equal structure, which splits the OFDM signal into multiple equal fragments to implement highly accurate Doppler shift estimation.

Inter-signal interference (ISI) caused by multipath effects cannot always be resolved by the short length of cyclic prefix (CP)\cite{7}. This often results in poor correlation between the data of the preamble sequence generating the ISI. Consequently, the accuracy of DFS estimation is adversely affected. Additionally, maintaining a certain length for the preamble sequence while increasing the number of repeated sequences can escalate algorithmic complexity, diminish real-time compensation performance, but broaden the range of DFS estimation. To address these issues, this paper proposes a novel approach in which three repeated sequences are combined to form a preamble sequence that occupies an OFDM symbol jointly. This method not only circumvents issues associated with ISI, but also enhances the feasibility and reliability of the DFS estimation algorithm.

Moreover, We propose a novel Doppler frequency offset algorithm that utilizes the radar-estimated DFS as a initial estimation, and design a preablem sequence for secondary estimation and compensation. Based on this, we design an adaptive DFS estimatior by incorporate the priori estimation results of DFS, which improve the performance of the system.

\section{SYSTEM MODEL}
\begin{figure}[htb]
	\centering
	\includegraphics[scale=0.55]{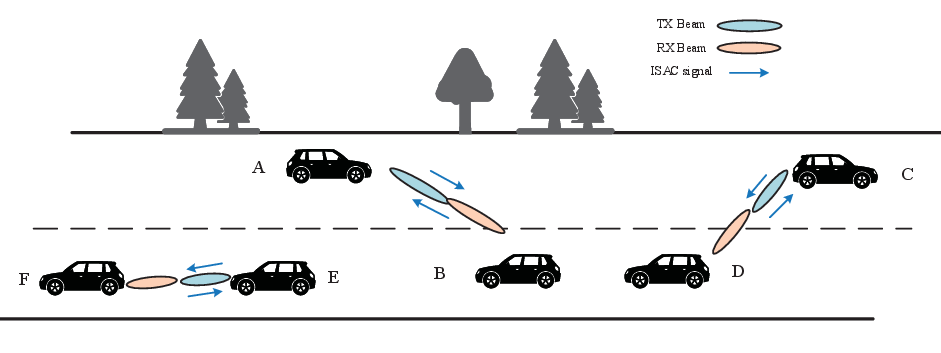}
	\caption{{A high dynamic mmWave vehicle network scenario. }}
	\label{FIG1}
\end{figure}
Fig. 1 shows a high dynamic mmWave vehicle network scenario. There are two types of communication links in V2V communication scenarios in Fig. 1, such as the communication link between vehicle E and vehicle F traveling in the same direction and the communication link between vehicle C and vehicle D traveling in the opposite direction. Assuming that vehicle C and vehicle D are both traveling at a maximum speed, then the relative speed between them is double their own speed. If the carrier frequency is relatively high, the receiver on vehicle C will experience a significant DFS, which can affect its performance. Furthermore, vehicles that frequently change lanes, shift, brake, or turn will cause greater variation in the expansion of DFS within the system\cite{3}. Hence, the high dynamicity of DFS in real-time necessitates precise compensation within the vehicular network scenario\cite{3}. Subsequently, we will introduce the V2V channel model in high dynamic scenarios.

Each vehicle is equipped with a uniform linear array (ULA) consisting of $P$ transmitting antennas and $P$ receiving antennas. In this paper, we focus on two vehicles driving in opposite directions, which is the foundation of multi-vehicles network.

In Fig. 1, vehicle A transmits a ISAC signal to vehicle B. Vehicle B obtains the communications data from vehicle A and processes the signal. At the same time, vehicle B receiver utilizes the received ISAC signal for sensing to obtain the DFS and performs the first compensation at the receiver. Then vehicle B makes a second fine estimate of DFS based on the preamble sequence.

Assuming that the baseband signal for single OFDM symbol transmitted from the transmitter of vehicle A is\cite{4}
\begin{align}
	x(n) = \frac{1}{N}\sum\limits_{k = 0}^{N - 1} {s(k) \cdot {e^{j2\pi nk/N}}} ,n = 0,...,N - 1,
\end{align}
where $N$ represents the number of subcarriers. $s(k)$ represents the constellation transmitted on the $k$th subcarrier. Assuming that there are ${L_{path}}$ paths in the V2V channel, due to the almost ignored impact of the millimeter wave signal on the receiver after reflection, the one-way channel response between a typical vehicle A and vehicle B in the millimeter wave system can be expressed as\cite{6}
\begin{equation}
h(t,\tau ) = \sum\limits_{l = 0}^{{L_{path}} - 1} {\tilde \alpha _l^r{e^{j2\pi {f_d}t}}{{\bf{a}}_{\bf{r}}}\left( {\theta _r^l} \right){\bf{a}}_{\bf{t}}^{\bf{H}}\left( {\theta _t^l} \right)} \delta (t - {\tau _l}),
\end{equation}
where $l$ is the $l$th propagation path. It's the direct path when $l=0$. $\tilde \alpha _l^r$ denotes the fading coefficient of the $l$th propagation path. ${\theta _r^l}$ and ${\theta _t^l}$ represent the angle of arrival (AoA) and angle of departure (AoD) of the $l$th path respectively. $\tau $ represents the time delay, and ${{f_d}}$ represents the DFS. ${{\bf{a}}_{\bf{r}}} \left( {\theta _r^l} \right)$ and ${\bf{a}}_{\bf{t}}^{\bf{H}}\left( {\theta _t^l} \right)$ denote the steering vectors of the transmitting and receiving antennas of the $l$th propagation path, and are expressed as\cite{6}
\begin{equation}
{{\bf{a}}_{\bf{t}}}\left( {\theta _t^l} \right) = \frac{1}{{\sqrt P }}\left[ \begin{array}{l}
	1\\
	\exp \left( {\frac{{ - j2\pi {\Delta _t}\sin {\theta _t}}}{\lambda }} \right)\\
	\exp \left( {\frac{{ - j2\pi  \cdot 2{\Delta _t}\sin {\theta _t}}}{\lambda }} \right)\\
	...\\
	\exp \left( {\frac{{ - j2\pi  \cdot (P - 1){\Delta _t}\sin {\theta _t}}}{\lambda }} \right)
\end{array} \right]^T,
\end{equation}
\begin{equation}
{{\bf{a}}_{\bf{r}}}\left( {\theta _r^l} \right) = \frac{1}{{\sqrt P }}\left[ \begin{array}{l}
	1\\
	\exp \left( {\frac{{ - j2\pi {\Delta _r}\sin {\theta _r}}}{\lambda }} \right)\\
	\exp \left( {\frac{{ - j2\pi  \cdot 2{\Delta _r}\sin {\theta _r}}}{\lambda }} \right)\\
	...\\
	\exp \left( {\frac{{ - j2\pi  \cdot (P - 1){\Delta _r}\sin {\theta _r}}}{\lambda }} \right)
\end{array} \right]^T,
\end{equation}
where the array element spacing is taken to be half a wavelength, i.e. ${\Delta _t} = {\Delta _r} = \frac{\lambda }{2}$, and the range of AoD and AoA at the receiver is $\left[ { - \frac{\pi }{2},\frac{\pi }{2}} \right]$. Therefore, affected by multipath effect, Doppler effect, etc., the time domain signal received at vehicle B can be denoted as\cite{6}
\begin{equation}
	y(n) = \sum\limits_{l = 0}^{{L_{path}} - 1} {\tilde \alpha _l^r{e^{j2\pi {f_d}n{T_s}}}{{\bf{a}}_{\bf{r}}}\left( {\theta _r^l} \right){\bf{a}}_{\bf{t}}^{\bf{H}}\left( {\theta _t^l} \right)} x(n - \tau_l ) + w(n),
\end{equation}
where $w(n)\sim CN(0,{\sigma ^2})$ additive white Gaussian noise. An effective beam alignment method can neglect the effect of scattering path on DFS expansion in millimeter wave systems.

The focus of this paper is on the compensation of the DFS, thus it is assumed that an effective beam alignment has been achieved and the signal is concentrated in the LOS path direction, meanwhile the interference from other paths is suppressed by the high-precision beam assignment\cite{beam}. Assuming that the weights of the beam assignment in the $\theta$-direction are ${\bf{c}}^{\bf{H}}(\theta _r^{})$ and ${\bf{b}}(\theta _t^{})$, the received signal after the ideal beam assignment is\cite{beam}
\begin{equation}
	\begin{aligned}
	Q(n) &= {\bf{c}}^{\bf{H}}(\theta _r^{})y(n){\bf{b}}(\theta _t^{})\\	
	&= T(n) + G(n) + h(n)
	\end{aligned},
\end{equation}
\begin{equation}
	T(n)=\sum\limits_{l,\theta _r^l = \theta _r^{},\theta _t^l = \theta _t^{}}^{} {\tilde \alpha _l^r{e^{j2\pi {f_d}n{T_s}}}} x(n - \tau ),
\end{equation}
\begin{align}
	G(n) &=\sum\limits_{l,\theta _r^l \ne \theta _r^{},\theta _t^l \ne \theta _t^{}}^{} {\tilde \alpha _l^r{e^{j2\pi {f_d}n{T_s}}}{\bf{c}}^{\bf{H}}(\varphi ){\bf{a}}_{\bf{t}}^{\bf{H}}\left( {\theta _r^l} \right){\bf{a}}_{\bf{t}}^{\bf{H}}\left( {\theta _t^l} \right)} \notag\\
	& \times x(n - {\tau _l}){\bf{b}}(\theta ),
\end{align}
\begin{align}
	h(n)={\bf{c}}^{\bf{H}}(\varphi )w(n){\bf{b}}(\theta ),
\end{align}
where $h(n)$ represents the noise term after beam assignment, $T(n)$ represents the received signal in the specified direction after beam assignment, $G(n)$ represents the interference term in the remaining direction after beam assignment.

\section{THE PROPOSED APPROACH}
Accurate and prompt estimation and compensation of DFS are crucial in high dynamic vehicle network scenarios. In this section, we propose an adaptive radar-assisted DFS estimation and compensation algorithm to fulfill the requirements of high accuracy and high reliability in vehicle communication.

\subsection{Radar Signal Processing}
When vehicle B receives the ISAC signal, the receiver of vehicle B processes the ISAC signal by the Two-Dimensional Fast Fourier Transform (2D-FFT) algorithm. We obtain the peak index in the range doppler map by the 2D-FFT algorithm\cite{strurm} as
\begin{equation}
\hat l = \mathop {\arg \max }\limits_l  |\frac{1}{M}\sum\limits_{n = 0}^{M - 1} {\exp (j2\pi {f_d}mT)\exp ( - j\frac{{2\pi ml}}{M})} |.
\end{equation}

When ${f_d}mT = \frac{{m\hat l}}{M}$, the DFS estimation result is obtained as ${f_d} = \frac{\hat l}{{MT}} = \frac{{v{f_c}}}{c}$, then the normalized DFS ${\varepsilon _v}$ is
\begin{equation}
{\varepsilon _v} = \frac{{v{f_c}}}{{c\Delta f}}.
\end{equation}
\subsection{Fine estimation and compensation of DFS based on preamble sequence}
The DFS obtained from the radar detection can be utilized as a coarse estimation result of DFS and is compensated to the received signal of vehicle B. Then the compensated received signal in the time domain can be expressed as
\begin{equation}
	\mathord{\buildrel{\lower3pt\hbox{$\scriptscriptstyle\frown$}} 
		\over y} (n) = y(n){e^{ - j2\pi {\varepsilon _v}k/N}}+w(n)',k=0,1,...,N-1.
\end{equation}
\begin{figure}[htb]
	\centering
	\includegraphics[scale=0.75]{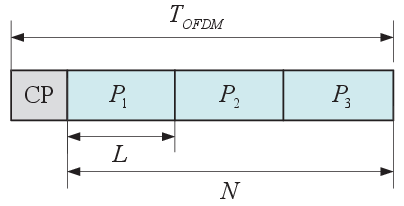}
	\caption{{The design of preamble sequence for the proposed DFS estimation algorithm. }}
	\label{FIG2}
\end{figure}

The DFS estimation results obtained by the radar are low in accuracy, which cannot meet the requirements of DFS estimation for vehicle communication scenarios. In this subsection, the fine estimation of DFS is based on the newly designed preamble sequence. Fig. 2 shows the preamble sequence design of the proposed DFS estimation algorithm, where the preamble sequence of duration ${T_{OFDM}}$ is located at the head of the frame structure and the lengths of the repeated sequence blocks ${P_1},{P_2},{P_3}$ are $L$.

We utilize two combinations of ${P_1}$ with ${P_2}$ and ${P_2}$ with ${P_3}$ to obtain the Doppler shift values for the second step of fine-grained estimation by the correlation operation of the MLE algorithm. Assuming that the 3 observed samples of the preamble sequence in the received signal are $\bar y = \{ \mathord{\buildrel{\lower3pt\hbox{$\scriptscriptstyle\frown$}} 
	\over y} (n)|n = 0,1,3\} $, the repeated sequence signal as ${P_i} = \{ iL,iL + 1,...,iL + L - 1\} ,i = 0,1,2$, define $f(\mathord{\buildrel{\lower3pt\hbox{$\scriptscriptstyle\frown$}} 
	\over y} |{\varepsilon _f})$ as the probability density function of the 3 observed samples in $\mathord{\buildrel{\lower3pt\hbox{$\scriptscriptstyle\frown$}} 
	\over y} (n)$  containing the DFS ${\varepsilon _f}$. 
According to $f(\mathord{\buildrel{\lower3pt\hbox{$\scriptscriptstyle\frown$}} 
	\over y} |{\varepsilon _f})$ the log likelihood function can be solved for $f(\mathord{\buildrel{\lower3pt\hbox{$\scriptscriptstyle\frown$}} 
	\over y} |{\varepsilon _f})$ for the maximum value\cite{ml}. The log likelihood function for $f(\mathord{\buildrel{\lower3pt\hbox{$\scriptscriptstyle\frown$}} 
	\over y} |{\varepsilon _f})$ is\cite{ml}
\begin{equation}
	\begin{aligned}
		\Lambda ({\varepsilon _f}) &= \log (f(\bar y|{\varepsilon _f}))\\
		&= \log \{ \prod\limits_{n \in {P_0}} {(f(\mathord{\buildrel{\lower3pt\hbox{$\scriptscriptstyle\frown$}} 
				\over y} (n),\mathord{\buildrel{\lower3pt\hbox{$\scriptscriptstyle\frown$}} 
				\over y} (n + L),\mathord{\buildrel{\lower3pt\hbox{$\scriptscriptstyle\frown$}} 
				\over y} (n + 2L)|{\varepsilon _f}))} \\
		&\times \prod\limits_{n \notin {P_0} \cup {P_1} \cup {P_2}} {(f(\mathord{\buildrel{\lower3pt\hbox{$\scriptscriptstyle\frown$}} 
				\over y} (n)|{\varepsilon _f}))} \} \\
		&= \log \{ \prod\limits_{n \in {P_0}} {\frac{{f(\bar I(n)|{\varepsilon _f})}}{{\prod\limits_{m = 0}^2 {f(\mathord{\buildrel{\lower3pt\hbox{$\scriptscriptstyle\frown$}} 
							\over y} (n + mL)|{\varepsilon _f})} }}\prod\limits_{m = 0}^2 {(f(\mathord{\buildrel{\lower3pt\hbox{$\scriptscriptstyle\frown$}} 
					\over y} (n)|{\varepsilon _f}))} } \} 
	\end{aligned},
\end{equation}
where $\bar I(n) = {[\mathord{\buildrel{\lower3pt\hbox{$\scriptscriptstyle\frown$}} 
		\over y} (n)\mathord{\buildrel{\lower3pt\hbox{$\scriptscriptstyle\frown$}} 
		\over y} (n + L)\mathord{\buildrel{\lower3pt\hbox{$\scriptscriptstyle\frown$}} 
		\over y} (n + 2L)]^T},n \in {P_i}$ obeying the 3-dimensional Gaussian distribution, thus the probability density function of $\bar I(n)$ is
\begin{equation}
f(\bar I(n)|{\varepsilon _f}) = \frac{{\exp ( - \bar I{{(n)}^H}{K^{ - 1}}\bar I(n))}}{{{\pi ^3}\det (K)}},
\end{equation}
where the matrix $K$ is a 3-dimensional matrix, ${K_{ij}} = E\{ \mathord{\buildrel{\lower3pt\hbox{$\scriptscriptstyle\frown$}} 
	\over y} (n + iL){\mathord{\buildrel{\lower3pt\hbox{$\scriptscriptstyle\frown$}} 
		\over y} ^*}(n + jL)\} ,n \in {P_0},i,j \in \{ 0,1,2\} $, suppose the variance of received signal is $\sigma _s^2$, the variance of noise is $\sigma _n^2$, the determinant of the matrix $K$ is $\det (K) = {\left( {\sigma _n^2} \right)^2}(3\sigma _s^2 + \sigma _n^2)$. $\mathord{\buildrel{\lower3pt\hbox{$\scriptscriptstyle\frown$}} 
	\over y} (n + mL),m = 0,1,2$ obeys a one-dimensional Gaussian distribution,and the probability density function of
\begin{equation}
	f(\mathord{\buildrel{\lower3pt\hbox{$\scriptscriptstyle\frown$}} 
		\over y} (n)|{\varepsilon _f}) = \frac{{\exp ( - \frac{{|\mathord{\buildrel{\lower3pt\hbox{$\scriptscriptstyle\frown$}} 
						\over y} (n){|^2}}}{{\sigma _s^2 + \sigma _n^2}})}}{{{\pi ^2}(\sigma _s^2 + \sigma _n^2)}}. 
\end{equation}	

The steps to simplify the log-likelihood function for $f(\mathord{\buildrel{\lower3pt\hbox{$\scriptscriptstyle\frown$}} 
	\over y} |{\varepsilon _f})$ are as follows,  
\begin{equation}
	\begin{aligned}
		\Lambda ({\varepsilon _f}) 
		&= \sum\limits_{n \in {P_0}} { - \bar I{{(n)}^H}{K^{ - 1}}\bar I(n)}  + \sum\limits_{i = 0}^2 {\frac{{|\mathord{\buildrel{\lower3pt\hbox{$\scriptscriptstyle\frown$}} 
						\over y} (n + iL){|^2}}}{{\sigma _s^2 + \sigma _n^2}}} \\
		&= 2\sum\limits_{m = 0}^2 {{\mathop{\rm Re}\nolimits} \{ {k^{ - m}}\varphi (i)\} }  - 2\rho \gamma ({\varepsilon _f})
	\end{aligned},
\end{equation}
\begin{equation}
\rho  = \frac{{\sigma _s^2}}{{\sigma _s^2 + \sigma _n^2}} ,
\end{equation}
\begin{equation}
\gamma ({\varepsilon _f}) = \sum\limits_{i = 0}^2 {\sum\limits_{n = 0}^{L - 1} {|\mathord{\buildrel{\lower3pt\hbox{$\scriptscriptstyle\frown$}} \over y} (n + iL){|^2}} },
\end{equation}
\begin{equation}
	\varphi (m) = \sum\limits_{i = 0}^{2 - m} {\sum\limits_{n = 0}^{L - 1} {\mathord{\buildrel{\lower3pt\hbox{$\scriptscriptstyle\frown$}} 
				\over y} (n + iL)\mathord{\buildrel{\lower3pt\hbox{$\scriptscriptstyle\frown$}} 
				\over y} (n + (i + m)L)} } ,
\end{equation}
\begin{equation}
	k = \exp (\frac{{ - j2\pi }}{3}).
\end{equation}

The maximum likelihood estimation result of ${\varepsilon _f}$ is ${\mathord{\buildrel{\lower3pt\hbox{$\scriptscriptstyle\frown$}} 
		\over \varepsilon } _f} = \arg \max \Lambda ({\varepsilon _f})$.
When $\frac{{\partial \Lambda ({\varepsilon _f})}}{{\partial {\varepsilon _f}}} = 0$, the maximum estimation result of the DFS is ${\mathord{\buildrel{\lower3pt\hbox{$\scriptscriptstyle\frown$}} 
		\over \varepsilon } _f}$. The first order derivative of the result can be represented as
\begin{equation}
	\begin{aligned}{l}
		\frac{{\partial \Lambda ({\varepsilon _f})}}{{\partial {\varepsilon _f}}} &= \frac{\partial }{{\partial {\varepsilon _f}}}\{ 2\sum\limits_{m = 0}^2 {{\mathop{\rm Re}\nolimits} \{ {k^{ - m}}\varphi (i)\} } \} \\
		&=  - \sum\limits_{m = 0}^2 {\frac{{4\pi m}}{3}|\varphi (m)|} \sin (\angle \varphi (m) + \frac{{2\pi {\varepsilon _f}m}}{3})\\
		&\approx  - \sum\limits_{m = 0}^2 {\frac{{4\pi m}}{3}|\varphi (m)|} (\angle \varphi (m) + \frac{{2\pi {\varepsilon _f}m}}{3})
	\end{aligned},
\end{equation}
then, the maximum likelihood estimation of the DFS is
\begin{equation}
	{\mathord{\buildrel{\lower3pt\hbox{$\scriptscriptstyle\frown$}} 
			\over \varepsilon } _f} =  - \frac{3}{{2\pi }}\frac{{|\varphi (1)|\angle \varphi (1) + 2|\varphi (2)|\angle \varphi (2)}}{{|\varphi (1)| + 4|\varphi (2)|}}.
\end{equation}

Thus, the complete DFS estimation is
\begin{equation}
	\mathord{\buildrel{\lower3pt\hbox{$\scriptscriptstyle\frown$}} 
		\over \varepsilon }  = {\mathord{\buildrel{\lower3pt\hbox{$\scriptscriptstyle\frown$}} 
			\over \varepsilon } _f} + {\mathord{\buildrel{\lower3pt\hbox{$\scriptscriptstyle\frown$}} 
			\over \varepsilon } _v}.
\end{equation}
\subsection{Cramer-Rao lower bound (CRLB) analysis of the proposed DFS estimation algorithm}
In this subsection, the CRLB of the proposed algorithm is analyzed, which measures the performance of unbiased estimation \cite{CRLB}. The received signal of the proposed algorithm is given as
\begin{equation}
\tilde y(k) = x(k){e^{j2\pi k\tilde \varepsilon /N}} + \tilde w(k),
\end{equation}
where $w(n)\sim CN(0,{\sigma_n ^2})$ additive white Gaussian noise. Observe the received signal with unknown parameters $\varepsilon $, and estimate $\varepsilon $, whose likelihood function is
\begin{equation}
	p(\tilde y;\varepsilon ) = \frac{{\exp [ - \frac{1}{{2\sigma _n^2}}\sum\limits_{k = 0}^{N - 1} {{{({\tilde y}(k) - x(k){e^{j2\pi k\tilde \varepsilon /N}})}^2}} ]}}{{{{\left( {2\pi \sigma _n^2} \right)}^{\frac{N}{2}}}}}.
\end{equation}

The log-likelihood function of $\tilde y(k)$ is
\begin{align}
	L(\tilde y;\varepsilon )& =  - \frac{N}{2}\ln \left( {2\pi \sigma _n^2} \right)\\& - \frac{1}{{2\sigma _n^2}}\sum\limits_{k = 0}^{N - 1} {{{({\tilde y}(k) - x(k){e^{j2\pi k\tilde \varepsilon /N}})}^2}} .
\end{align}

According to the method of calculating the CRLB in the literature \cite{CRLB1}, we can know that $Var(\mathord{\buildrel{\lower3pt\hbox{$\scriptscriptstyle\frown$}} 
	\over \varepsilon } ) \ge \frac{1}{{I(\varepsilon )}}$ and the fisher matrix is 
\begin{equation}
	I(\varepsilon ) =  - E\left[ {\frac{{\partial L(\tilde y;\varepsilon )}}{{\partial {\varepsilon ^2}}}} \right],
\end{equation}
and the CRLB of $\varepsilon $ is
\begin{equation}
	\begin{array}{l}
		Var(\mathord{\buildrel{\lower3pt\hbox{$\scriptscriptstyle\frown$}} 
			\over \varepsilon } ) \ge \frac{3}{{2{\pi ^2}LSNR}}
	\end{array}.
\end{equation}

It can be seen that the CRLB of $\varepsilon $ is related to the value of SNR taken when the length $L$ of the repeated sequence is fixed. With the increase of SNR, the CRLB of $\varepsilon $ will decrease.

\subsection{Sensing assisted adaptive estimation and compensation algorithm of DFS}
\begin{figure}[htb]
	\centering
	\includegraphics[scale=0.53]{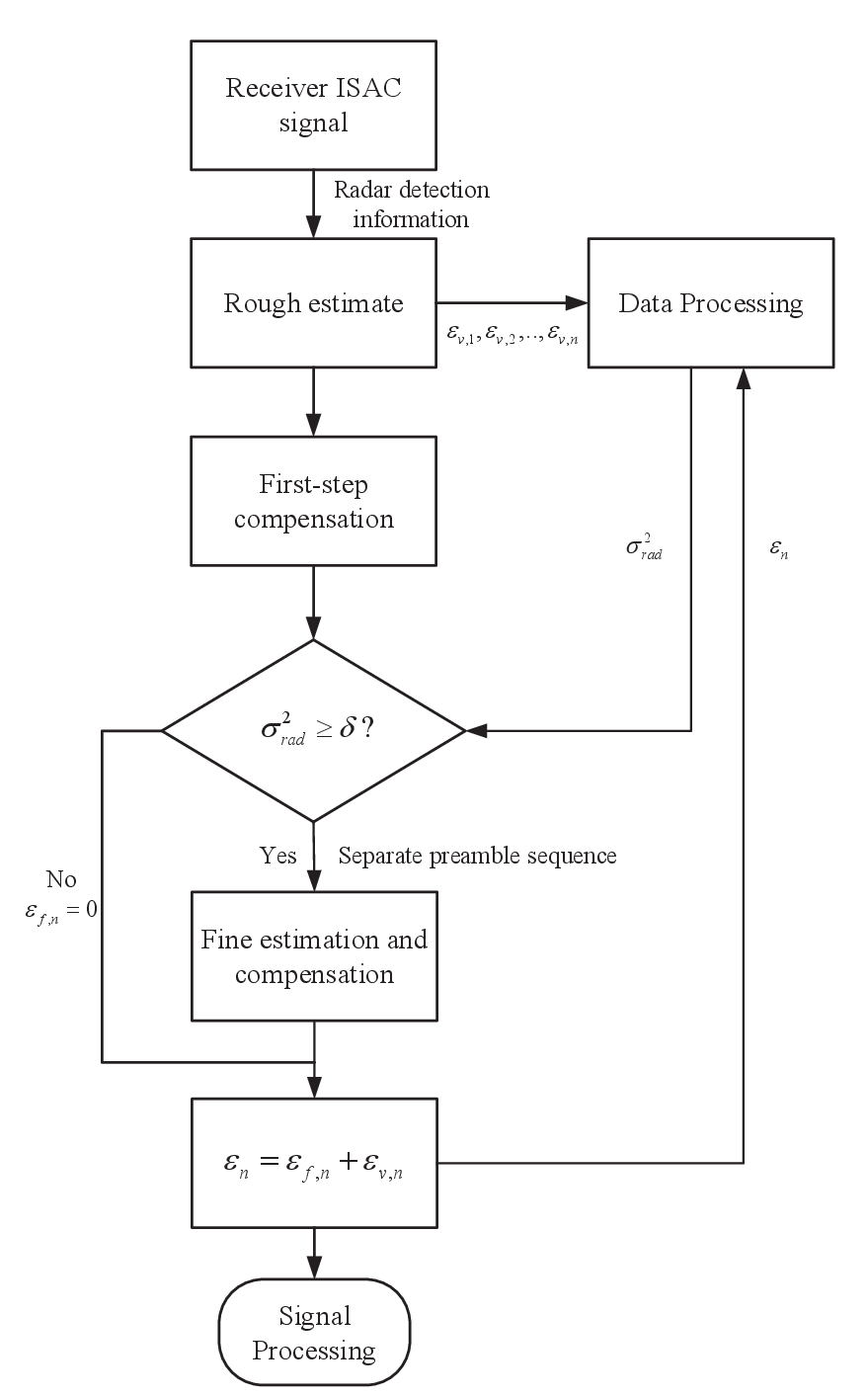}
	\caption{{Sensing-assisted adaptive estimation and compensation algorithm of DFS. }}
	\label{FIG3}
\end{figure}
The accuracy of the estimation results of DFS obtained from radar detection is relatively low, and the fine estimation based on the preamble sequence will increase the complexity of the algorithm. Therefore, we propose an adaptive DFS estimation compensation algorithm, which will improve the estimation efficiency of the system. Fig. 3 shows the flowchart of the proposed sensing-assisted adaptive DFS estimation and compensation algorithm.

The priori estimation results of DFS from radar detection ${\varepsilon _{v,1}},{\varepsilon _{v,2}},..,{\varepsilon _{v,n}}$ are introduced, and the variance $\sigma _{rad}^2 = \frac{1}{n}\sum\limits_{n = 1}^n {{{({{\bar \varepsilon }_h} - {\varepsilon _{v,n}})}^2}} $ of the priori estimation datas of  DFS is calculated, which is employed to characterize current change rate of DFS.
Meanwhile, we establish a threshold $\delta $ to judge the degree of fluctuation of DFS. When the rate of change of the DFS exceeds the threshold, that is $\sigma _{rad}^2 \ge \delta $, the DFS is determined to be high-dynamic ,and then fine estimation and compensation are performed. When the fluctuation of DFS is not obvious, that is $\sigma _{rad}^2 < \delta $, only primary estimation and compensation are required to fulfill the requirement of DFS compensation. By the above-mentioned algorithm of adaptive DFS compensation, the complexity of the proposed algorithm is properly reduced, and the real-time and accurate DFS compensation will be realized.  

\section{NUMERICAL RESULTS}
In order to verify the effectiveness of the sensing-assisted adaptive DFS estimation and compensation algorithm proposed in Section III, the simulation results are presented in this section. And we compare the sensing-assisted DFS estimation algorithm with  various conventional algorithms and analyze the rationality and advantages of the proposed algorithm. Finally, we evaluate the improvement of communication performance by the proposed algorithm.

In the simulation experiments, we utilize the MATLAB simulation platform to implement the complete and effective communication link, as well as the channel modeling simulation of the vehicular commnunication scenario. Based on the parameter design of typical OFDM communication link systems with reference, the main parameters of the simulation platform are set as shown in the TABLE I\cite{3GPP 38.901}.

\begin{table}[h]
	\caption{\label{sys_para}Simulation parameters}
	\begin{center}
		\begin{tabular}{l l}
			\hline
			\hline
			
			{Parameter} & {Setting and Description}\\
			
			\hline
			
			FFT size & 128\\
			Cyclic prefix length & $16$ \\
			Preamble sequence length & $64$ \\
			Modulation & 16QAM\\
			Normalized maximum DFS & $0.25$ \\
			Channel Environment & AWGN\\
			
			\hline
			\hline
		\end{tabular}
	\end{center}
	\label{table_1}
\end{table}

The effectiveness of the proposed algorithm is verified by analyzing and comparing the Mean Squared Error (MSE) and Bit Error Rate (BER) performance of the traditional algorithm and the proposed algorithm. Fig. 4 shows the MSE versus SNR curves of CP based estimator (CPBE)\cite{CPBE}, Pilot Symbol Aided (PSA)\cite{PSA}, Moose\cite{Moose} and CRLB. Compared with PSA algorithm, Moose algorithm and the proposed algorithm, the MSE performance of CPBE algorithm is relatively poor. When the system SNR is higher than 5dB, the MSE performance of the Moose algorithm is much better than that of the PSA algorithm. Obviously, the algorithm proposed in this paper has the best MSE performance and is closest to the CRLB curve, relatively. It can be seen that when the SNR of the system is higher, the MSE curve of the proposed algorithm is closer to the CRLB curve ,thus the proposed algorithm is unbiased estimation.

\begin{figure}[htb]
	\centering
	\includegraphics[scale=0.52]{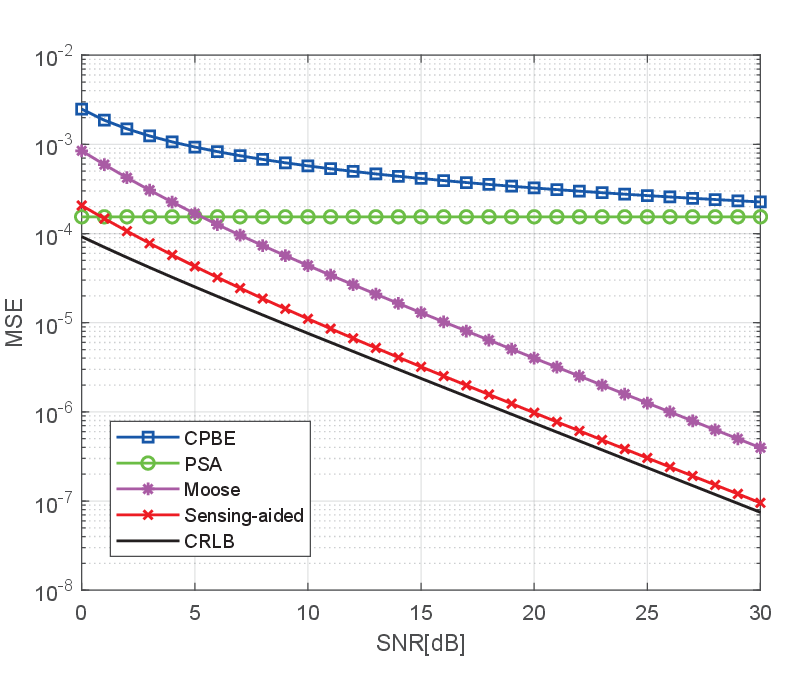}
	\caption{{The system MSE versus SNR under different DFS estimaition algorithms. }}
	\label{FIG4}
\end{figure}
\begin{figure}[htb]
	\centering
	\includegraphics[scale=0.52]{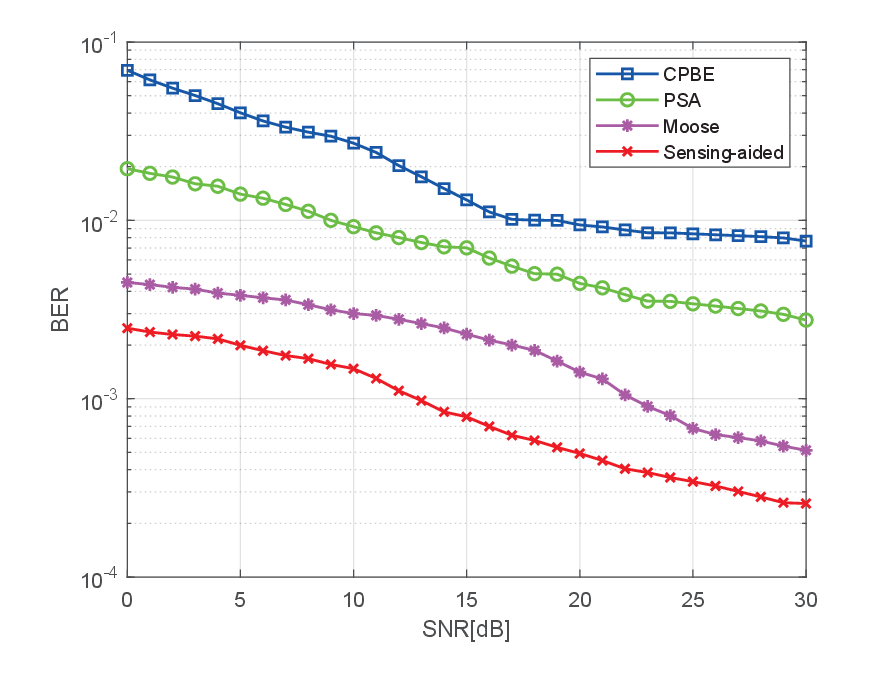}
	\caption{{The system BER versus SNR under different DFS estimaition algorithms. }}
	\label{FIG4}
\end{figure}
Fig. 5 shows the system BER versus SNR under different DFS estimaition algorithms. Under the conditions of the same SNR, the ranking of BER performance of the algorithms from best to worst is CPBE algorithm, PSA algorithm, Moose algorithm, and sensing-assisted DFS algorithm. In summary, both of the BER performance and MSE performance of the proposed algorithm in this paper are excellent, relatively.

Fig. 6 and Fig. 7 show the proposed algorithm performance when the DFS fluctuates highly under the SNR of 3 dB and 20 dB, respectively. In order to simulate the variation of DFS in high mobility vehicle communication scenarios, we set the ideal normalized DFS as a random number fluctuating between 0.1 and 0.25. It can be seen that when the SNR is 20 dB, the performance of the proposed DFS estimation algorithm is better than the SNR is 3 dB. The proposed algorithm can quickly and accurately estimate changes of high mobile DFS, which is suitable for high dynamic vehicular communication scenarios. Fig. 8 shows the constellation diagram of the received signal before and after compensating for DFS. Without compensation, the received signal cannot be modulated correctly, resulting in distorted constellation diagram. By using the proposed algorithm to compensate for the DFS, the constellation diagram of the received signal can restore.

\begin{figure}[htb]
	\centering
	\includegraphics[scale=0.52]{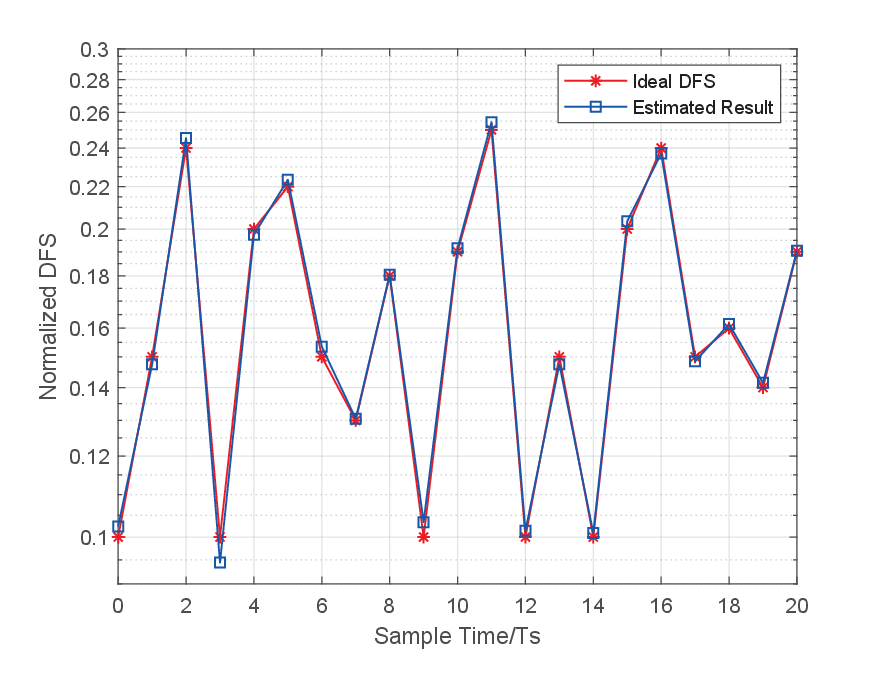}
	\caption{{High dynamic DFS estimation performance of the proposed algorithm ($20$ dB). }}
	\label{FIG4}
\end{figure}
\begin{figure}[htb]
	\centering
	\includegraphics[scale=0.52]{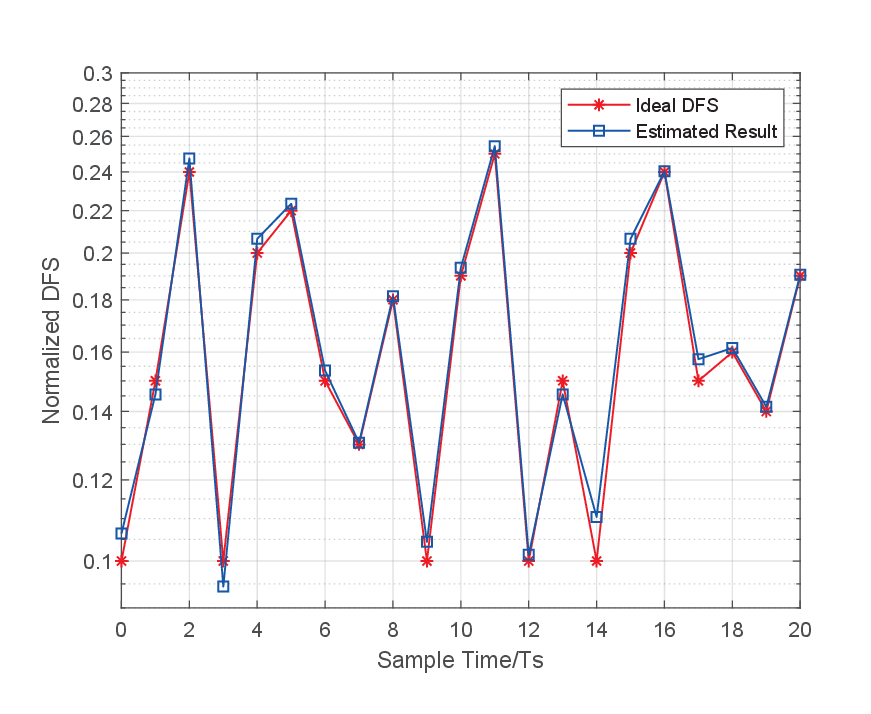}
	\caption{{High dynamic DFS estimation performance of the proposed algorithm ($3$ dB). }}
	\label{FIG4}
\end{figure}

\begin{figure}[htb]
	\centering
	\includegraphics[scale=0.35]{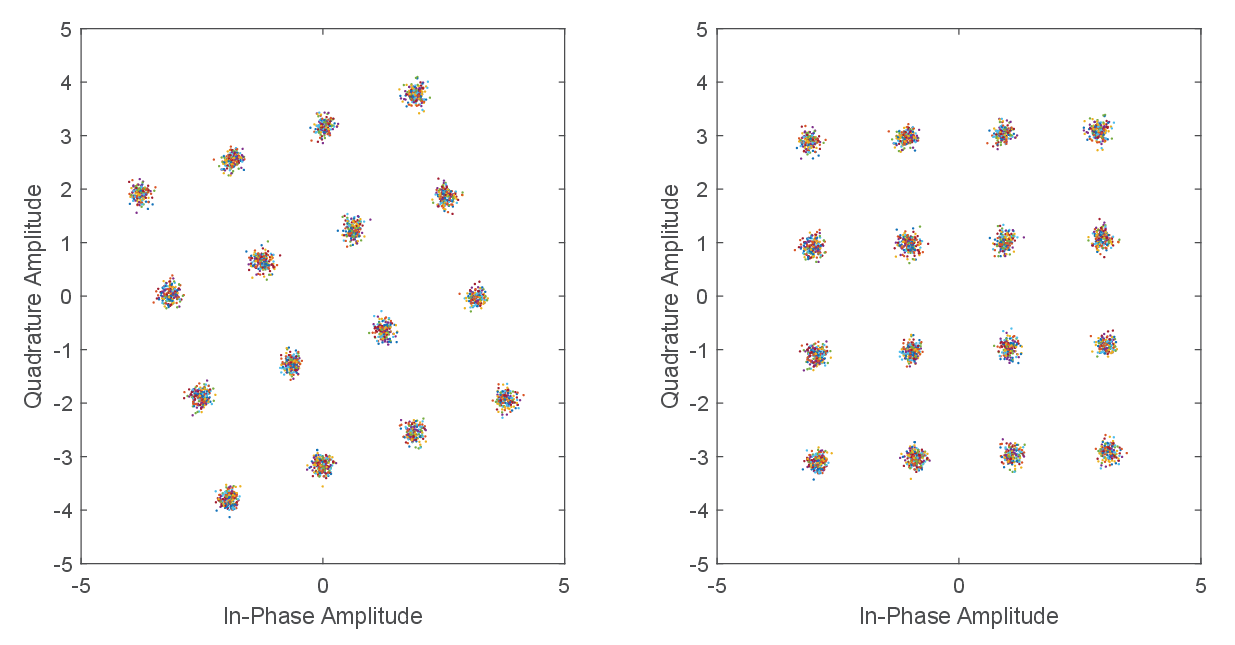}
	\caption{{Constellation diagram of the received signal before and after compensating for DFS. }}
	\label{FIG4}
\end{figure}

\section{Conclusion}
In this paper, we propose a sensing-assisted adaptive estimation algorithm of DFS.  Firstly, the DFS is coarsely estimated and compensated using radar detection.  Then, the designed preamble sequence is used to accurately estimate and compensate DFS.  In addition, an adaptive DFS estimator is designed to reduce the computational complexity in Section III.  The simulation results indicate that the sensing-assisted DFS estimation algorithm proposed in this paper can be capable of completing DFS estimation and compensation accurately and rapidly.  And it is suitable for high dynamic vehicular communication scenarios.

\section*{Acknowledgment}
This work was supported in part by the Beijing Municipal Natural Science Foundation under Grant 4212005, in part by the National Natural Science Foundation of China (NSFC) under Grant 92267202, 62271076, U21B2014, and in part by the National Key Research and Development Program of China under Grant 2020YFA0711302.


\begin{thebibliography}{00}
	
\bibitem{liu fan} S. Lu, F. Liu, Li. Y, et al.,“Integrated sensing and communications: Recent advances and ten open challenges,” {\em arXiv preprint arXiv}:2305.00179, 2023.	 

\bibitem{2} Z. Du, F. Liu, et al.,“Towards ISAC-empowered vehicular networks: Framework, advances, and opportunities,” {\em arXiv preprint arXiv}:2305.00681, 2023.

\bibitem{3} Jameel, Furqan, et al.,“Propagation channels for mmWave vehicular communications: State-of-the-art and future research directions,” {\em IEEE Wireless Communications.}, vol. 26, no. 1, pp. 144-150, 2018.

\bibitem{4} Ozkaptan C D, et al.,“OFDM pilot-based radar for joint vehicular communication and radar systems,” {\em 2018 IEEE Vehicular Networking Conference (VNC).}, 2018, pp. 1-8.

\bibitem{5} Nyongesa F C, Olwal T O, et al.,“Doppler shift estimation and compensation in high speed MIMO-OFDM VANETs,” {\em Journal of computing and information technology.}, vol. 26, no. 3, pp. 141-156, 2018.

\bibitem{ma} J. Ma, Qingpeng, et al., “Training Sequence Based Doppler Shift Estimation for Vehicular Communication,” {\em 2020 IEEE Wireless Communications and Networking Conference (WCNC).}, 2020, pp. 1-6.

\bibitem{6} Zhang C, Wang G, Jia M, et al., “Doppler shift estimation for millimeter-wave communication systems on high-speed railways,” {\em IEEE Access.}, vol. 7, pp. 40454--40462, 2018.

\bibitem{7} Zhang T, Xia X G.,“OFDM synthetic aperture radar imaging with sufficient cyclic prefix,” {\em IEEE Transactions on Geoscience and Remote Sensing.}, vol. 53,no. 1, pp. 394-404, 2014.

\bibitem{beam}W. Guo, W. Zhang, P. Mu, and F. Gao,“High-mobility OFDM downlink transmission with large-scale antenna array,”  {\em IEEE Trans. Veh. Technol.}, vol. 66, no. 9, pp. 8600-8604, Sept. 2017.

\bibitem{strurm} C. Sturm and W. Wiesbeck, “Waveform design and signal processing aspects for fusion of wireless communications and radar sensing,” {\em Proc. IEEE}, vol. 99, no. 7, pp. 1236-1259, Jul. 2011.

\bibitem{ml} M. Morelli, L. Marchetti and M. Moretti, "Maximum Likelihood Frequency Estimation and Preamble Identification in OFDMA-based WiMAX Systems," {\em IEEE Transactions on Wireless Communications,}  vol. 13, no. 3, pp. 1582-1592, March 2014.

\bibitem{CRLB} S. M. Kay, “Fundamentals of statistical signal processing: estimation theory,” {\em Proc. Prentice-Hall}, 1993.

\bibitem{CRLB1} Zhang Q, Sun H, Feng Z, et al., “Data-aided Doppler frequency shift estimation and compensation for UAVs,” {\em IEEE Internet of Things Journal,} vol. 7, no. 1, pp. 400-415, 2019. 

\bibitem{3GPP 38.901} 3GPP, TR 38.901 (V17.0.0), “Study on channel model for frequencies from 0.5 to 100 GHz,” Mar. 2022.

\bibitem{CPBE} J. J. van de Beek, M. Sandell, and P. O. Borjesson, “ML estimation of time and frequency offset in OFDM systems,” {\em IEEE Trans. Signal Process.}, vol. 45, no. 7, pp. 1800–1805, Jul. 1997.

\bibitem{PSA} Q. Shenping, Y. Changchuan, L. Jianfeng, and Y. Guangxin, “Pilot symbol-aided frequency offset estimation and correction for OFDM system,” {\em IEEE PIMRC.}, Sep. 2003, pp. 593–596.

\bibitem{Moose} E. S. Kang, H. Hwang, and D. S. Han, “A fine carrier recovery algorithm robust to Doppler shift for OFDM systems,” {\em IEEE Trans. Consum. Electron.}, vol. 56, no. 3, pp. 1218–1222, Aug. 2010.

\end{thebibliography}
\end{document}